\documentclass[conference]{IEEEtran}
\IEEEoverridecommandlockouts
\usepackage{cite}
\usepackage{amsmath,amssymb,amsfonts}
\usepackage{algorithmic}
\usepackage{graphicx}
\usepackage{textcomp}
\usepackage{xcolor}
\def\BibTeX{{\rm B\kern-.05em{\sc i\kern-.025em b}\kern-.08em
    T\kern-.1667em\lower.7ex\hbox{E}\kern-.125emX}}

\usepackage[hyphens]{url}

\newcounter{rq}[section]
\newenvironment{rqenv}[1][]{\refstepcounter{rq}
   \textbf{RQ\therq: #1}}{}

\begin{document}

\title{\textsc{Convo}: What does conversational \\ programming need? An exploration of machine learning interface design}

\author{
\IEEEauthorblockN{Jessica Van Brummelen}
\IEEEauthorblockA{jess@csail.mit.edu \\ MIT}
\and
\IEEEauthorblockN{Kevin Weng}
\IEEEauthorblockA{kweng@mit.edu \\ MIT}
\and
\IEEEauthorblockN{Phoebe Lin}
\IEEEauthorblockA{phoebelin@gsd.harvard.edu \\ Harvard Graduate School of Design}
\and
\IEEEauthorblockN{Catherine Yeo}
\IEEEauthorblockA{cyeo@college.harvard.edu \\ Harvard University}
}

\maketitle

\begin{abstract}
Vast improvements in natural language understanding and speech recognition have paved the way for conversational interaction with computers. While conversational agents have often been used for %
short goal-oriented dialog, we know little about agents for developing computer programs. To explore the utility of natural language for programming, we conducted a study ($n$=45) comparing different input methods to a conversational programming system we developed. Participants completed novice and advanced tasks using voice-based, text-based, and voice-or-text-based systems. We found that users appreciated aspects of each system (e.g., voice-input efficiency, text-input precision) and that novice users were more optimistic about programming using voice-input than advanced users. Our results show that future conversational programming tools should be tailored to users’ programming experience and allow users to choose their preferred input mode. To reduce cognitive load, future interfaces can incorporate visualizations and possess custom natural language understanding and speech recognition models for programming.

\end{abstract}

\begin{IEEEkeywords}
conversational programming,
conversational AI,
interaction paradigms,
voice interfaces,
accessibility,
education,
natural language processing,
human-computer interaction
\end{IEEEkeywords}

\section{Introduction}\label{sec:intro} %

With recent major advances in automatic speech recognition (ASR) and natural language processing (NLP)%
\cite{asr-attention,deep-speech-2-end-to-end,BERT}, interacting with technology has become as easy as having a conversation. %
Conversational agents have proliferated such that %
it would be shocking for a smartphone not to be able to transcribe %
and take action based on something someone said. Programmers have begun to automate simple, few-turn tasks using conversational artificial intelligence (AI), like turning on lights, as well as longer, more complex tasks, such as conversing with hairdressers to book clients' appointments \cite{google-duplex}.

\begin{figure*}[ht]
    \includegraphics[width=\linewidth]{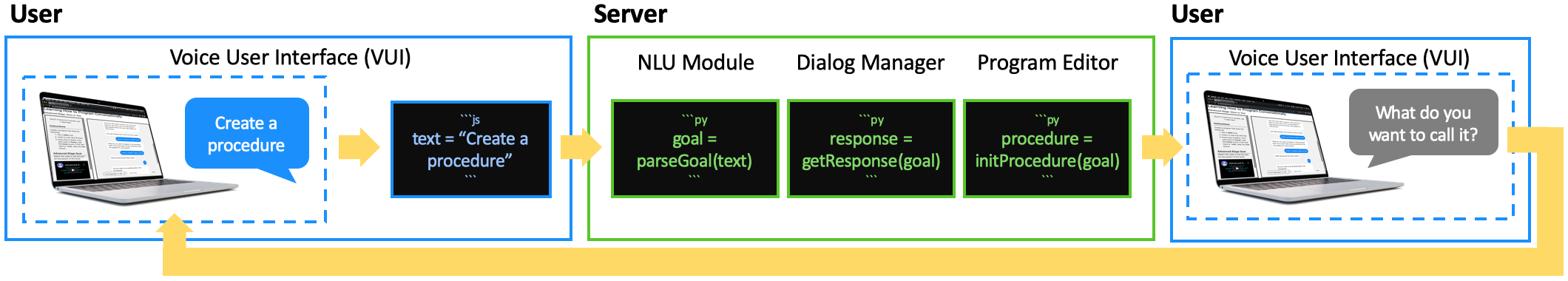}
    \caption{\textsc{Convo}'s architecture. The VUI passes spoken or typed input to the NLU module, which recognizes the intent. Given the intent, the DM might pass a response like, ``What do you want to call it?'', to the VUI to speak, or pass goals like, \texttt{create procedure}, to the PE.}
    \label{fig:convo-architecture}
\end{figure*}

With such advances, this technology is positioned to be leveraged in other spaces too. Specifically, it can increase technology accessibility through question-answering (QA), providing alternative input methods, and not requiring reading/writing skills for interaction. %
Computer programming could especially benefit from these advantages. Engaging learners in straightforward conversations without syntax requirements, could lower the barrier to entry to programming and provide efficient, alternative input methods.  Nonetheless, there are limitations to this technology. For example, ASR can be frustrating, especially for those with nontraditional accents, and NL is innately ambiguous, which could produce additional errors (e.g., is the meaning of ``say variable var'', \textit{say the value of variable var} or \textit{say the words ``variable var''}?).

Currently, little is known about the suitability of conversational agents for lowering the barrier to entry to programming. There has been some work in single-turn program synthesis, in which a NL utterance is converted to a program without conversation \cite{classic-program-synth-paper,program-synth-neurips}; syntax- or keyword-dependent, voice-based programming \cite{speech-spoken-java-vlhcc,voice-coding-nature}; conversational agents for controlling specific systems, such as Arduinos \cite{heyteddy}; and for learning linear tasks, such as sending emails \cite{LIA-SUGILITE}. However, optimal interaction paradigms for NL, conversational systems remain largely unstudied. We do not know whether it would be best for end-users to conversationally interact through speech, text, or a combination of multiple modalities. We also do not know how programming conversationally affects cognitive load and learning for novice programmers.

This paper will address questions about the usability, feasibility, and cognitive load of a conversational programming system. We completed a study ($n$=45) with our voice- and text-based conversational programming tool, in which users completed novice and advanced programming tasks using only voice-input, only text-input, and voice-or-text systems. They then answered Likert-scale and short answer questions about usability, satisfaction, preference and overall experience. Furthermore, we collected cognitive load indicator data, such as time to completion, %
number of system resets, lengths of utterances, and number of times users asked for help. We analyzed these data through quantitative and thematic analyses. The results will inform the development of future interactive machine learning (ML) systems, especially conversational programming agents, and begin to address the following research questions:

\begin{rqenv}\label{RQ:input}
What is the preferred input modality for a conversational tool? Would multimodal input be useful? 
\end{rqenv}

\begin{rqenv}\label{RQ:cogload}
How do input modalities affect cognitive load?%
\end{rqenv}

\begin{rqenv}\label{RQ:novadv}
Do novice and advanced programmers' conversational programming preferences differ?%
\end{rqenv}

\begin{rqenv}\label{RQ:asrnl}
Are current ASR and NLU technologies adequate for conversational coding?
\end{rqenv}

\begin{rqenv}\label{RQ:constrainednl}
Is it better to have constrained or unconstrained NL?
\end{rqenv}

\begin{rqenv}\label{RQ:edu}
Can conversational programming teach computational thinking skills? How does it compare to visual programming? %
\end{rqenv}

\begin{rqenv}\label{RQ:compaction}
How can a conversational programming system facilitate project creation and computational action \cite{computational-action}?
\end{rqenv}

This study is the first in a series aimed to address these questions. It addresses voice- and text-input preferences (RQ\ref{RQ:input}), cognitive load effects of voice- and text-input (RQ\ref{RQ:cogload}), advanced and novice user preferences (RQ\ref{RQ:novadv}), and out-of-the-box ASR and constrained NLU effectiveness (RQ\ref{RQ:asrnl} and RQ\ref{RQ:constrainednl}). Future studies with updated systems (e.g., unconstrained NLU system, system with visualizations) will address remaining questions. Ultimately, our goal is to leverage state-of-the-art ML technologies to empower learners to develop programs and solve problems in their communities. To do so, we will explore conversational AI design spaces with respect to lowering the barrier to entry to programming.  %

This paper presents the following main contributions:
\begin{enumerate}
    \item A formative study ($n$=45) examining cognitive load, input modalities, and advanced and novice programmers' performance with a conversational programming agent.
    \item The system design of a conversational programming agent, \textsc{Convo}.
    \item Design considerations for future systems based on quantitative and thematic analyses of user feedback.
\end{enumerate}

\section{Background and Related Work}\label{sec:bg} %
Our conversational programming system, \textsc{Convo}, draws on a number of research areas, including  conversational AI, voice-first human-computer interaction (HCI), natural language programming, %
and accessible programming.

\subsection{Conversational AI, Voice-First HCI, and NL Programming}\label{sec:bg-convai}
To create an effective conversational programming system, we utilized conversational and voice-first design principles. Conversational AI design literature often references Grice's ``conversational maxims'' \cite{grice}, which can be decomposed into \textit{concise}, \textit{correct}, \textit{relevant}, and \textit{natural} principles \cite{sm-jvanb}. %
For example, %
to adhere to the natural principle, \textsc{Convo} says ``The name, \texttt{name}, has already been used'' instead of ``Class \texttt{name} already exists''.

Another important principle is NLU flexibility. Design guides suggest that conversational systems should understand synonyms, over-answering and subtextual meaning \cite{voice-amazon,conversation-google}.
However, since we wanted to study the effects of constraining NL in a conversational programming system (RQ\ref{RQ:constrainednl}), we restricted \textsc{Convo} to understanding particular NL phrases, like ``create a variable'', so that we can compare it to an unconstrained NL system in future studies. We hypothesize that constraining NL input will reduce potential for ambiguity and increase the likelihood of \textsc{Convo} understanding users better; though at the same time, reducing the total number of phrases \textsc{Convo} understands could force users to think about precise commands, thereby increasing their cognitive loads. 

Previous studies support this hypothesis and illustrate how unconstrained NL results in ambiguity and can cause mismatch in human perception and reality of the system's understanding  \cite{hci-handbook,unconstrained-nl-bad}. Other research suggests, however, that such ambiguities can be eliminated through conversational QA and programming by demonstration \cite{codi-tina-quach,pumice-lia}. To the authors' knowledge, however, there have been no studies directly comparing constrained and unconstrained NL programming systems with such ambiguity reduction methods. In this study, we investigate users' perception of constrained NL with \textsc{Convo}.%

Historically, voice-based programming systems have been developed for advanced programmers using syntactically-constrained vocabularies (largely due to limited ASR and NLU technology) \cite{voicegrip-good-review-sec,vocalide,oneofthefirst-voice-programming,voice-coding-nature}. Recently, however, a number of voice-based, NL systems have appeared. \textit{TurtleTalk}, for instance, allows children to program the movement of a turtle in a speech-based video game environment \cite{turtletalk}. Other systems allow for mechatronic system control or database queries \cite{robot-nl-voice-programming,heyteddy,sql-nl-voice-programming}. Nonetheless, each of these systems are limited in scope. %
Though domain-specific ASR systems constrain the vocabulary space which generally results in better speech recognition, general-purpose ASR systems can provide the flexibility learning technology usually requires.
This research aims to determine the feasibility of a general-purpose voice-or-text NL programming system using current state-of-the-art ASR from Google Cloud Speech-to-Text API \cite{google-cloud-speech-to-text}.

\subsection{Accessibility and Cognitive Load}\label{sec:bg-accessibility}
Conversational coding systems have the potential to increase programming accessibility in three main areas: (1) for those with visual- and motor-impairments, (2) for those who are unable to read or type, and (3) for those with little prior programming knowledge. %
Particularly salient design principles for accessible auditory programming environments include describing code, not syntax; expressing \textit{logical context} (e.g., nested loop structure) over \textit{spatial context} (e.g., line 484); providing localization, querying and navigation cues; and intentionally addressing voice-based ambiguity, like homonyms \cite{blind-sighted-comparison,exploratory-study-blind-programmers,sodbeans-programming-blind-stefik,auditory-hci-stefik,non-sighted-programming-thesis-stefik,ast-programming-accessible-schanzer}. These principles guided the development of \textsc{Convo}.

Despite these principles, voice-based programming systems can still incite high cognitive load, especially when the vocabulary or grammar deviates greatly from natural language \cite{voicegrip-good-review-sec,blind-sighted-comparison}. %
Cognitive theory suggests that auditory and visual information are processed via separate channels, and that channels have limited capacity \cite{mayer2003promise}. Thus, a voice-based system without any scaffolding may overload the auditory channel, but a combination of the two (e.g., a voice-or-text-based system) may overcome the challenges of high cognitive load \cite{alex259193}. Recent research also shows that student learning is improved when students have verbal interaction back to a conversational system \cite{lepadatu2012use}.
\section{System Design}\label{sec:sys-design}
\textsc{Convo} is a voice-based system allowing users to develop computer programs by conversing in natural language with a conversational agent. For the user study, the system was designed to support both voice- and text-based conversations. Currently, the system supports three main tasks---(1) program creation, (2) program editing, and (3) system feedback---through natural language. %
This is illustrated in the following system walk-through.

\subsection{System Walk-through}
Currently, \textsc{Convo} is a constrained NL-based programming system in which commands have to be stated exactly for the system to understand. However, we are developing a less constrained version for future studies to analyze NL constraint effects on cognitive load. Here, we illustrate an example scenario and conversation using the current system with Lisa, a user, who wants to make a game for her little brother Chris to help him learn what sounds animals make. 

\textbf{Constrained NL System (Current).} \textit{Lisa starts up Convo and says ``Hey Convo, I want to make a game.'' Since the system doesn't recognize the word ``game'', it responds with ``I didn't understand what you want to do. You can start making a program by saying `create a program'." Lisa says, ``Okay, \textbf{create a program}.'' Convo responds by asking her, ``What do you want to name the program?''. Lisa replies, ``\textnormal{Animal Sounds}.'' She can now add actions to the program. Lisa wants to make a loop but does not know how. She sees an example phrase in the sidebar -- ``\textbf{Create a loop}'' -- and tries it out. The system asks for the halting condition, to which Lisa responds, ``\textbf{Until} I say `stop'." Now inside the loop, Lisa tells Convo to ask for user input by saying, ``\textbf{Get user input} and \textbf{save} it as \textnormal{animal}.'' Next, Lisa makes conditionals for %
dog, cat, horse, and cow sounds. For the dog sound, Lisa says ``\textbf{If} \textnormal{animal} is dog, \textbf{play} the dog \textbf{sound}.'' After going through each animal sound, Lisa closes the loop by saying, ``\textbf{Close loop}'', and tells Convo she is finished by stating ``\textbf{Done}.'' Finally, Lisa and her little brother Chris run the program by telling Convo to ``\textbf{Play} \textnormal{Animal Sounds}.''}

As illustrated by this scenario, \textsc{Convo} only recognizes exact commands like ``create a program'' or ``make a program,'' limiting the possible conversations that users can have with \textsc{Convo}. In addition, \textsc{Convo} has a constrained system for providing feedback or assistance to the user. Below, we illustrate the same scenario with a less constrained NL system. 

\textbf{Unconstrained NL System.} \textit{Lisa starts up Convo and says ``Hey Convo, can I \textbf{make a game} called \textnormal{Animal Sounds}?'' Convo does not know what a game is but is able to ask Lisa, ``What's a game? Is it like a procedure or a variable?". Convo responds ``It is a procedure." With that, Convo is able to equate ``making a game'' to making a procedure. In addition, Convo also recognizes that ``\textnormal{Animal Sounds}'' is the name that Lisa wants to use.
Convo proceeds to create a program called ``\textnormal{Animal Sounds}''. Lisa wants to make a loop, but she has never done it before with Convo. Instead of needing to consult documentation, she directly asks ``Convo, \textbf{how} do I make a loop?''. Convo directs her through the process by responding, ``First, you need to have a stopping condition. What do you want it to be?''. Lisa says, ``\textbf{Until} I say `stop'," and proceeds to add actions to the loop. She says ``\textbf{Set} \textnormal{animal} to user input''. Convo detects potential ambiguity and asks Lisa, ``Do you mean \textbf{get user input} and \textbf{set} `\textnormal{animal}' to the value of the input, or \textbf{set} `\textnormal{animal}' to the words, `user input'?'' Lisa indicates the former and proceeds to create the rest of the program.}

An unconstrained version of \textsc{Convo} would be able to recognize and detect user intents from a larger variety of utterances. Because ambiguity exists in natural conversations, the unconstrained version would also need to detect ambiguity and ask for clarification. %
By implementing such a system, we would be able to determine the suitability of both constrained and unconstrained conversational NL for programming.

\subsection{Technical Implementation}
\textsc{Convo} consists of four modules: the voice-user interface (VUI), natural language understanding (NLU) module, dialog manager (DM), and program editor (PE), as shown in Fig. \ref{fig:convo-architecture}.

Users interact with the system through the VUI. The VUI receives and transcribes voice input into text and sends it to the next module. The ASR is handled by Google's Cloud Speech-To-Text API \cite{google-cloud-speech-to-text} service. The transcribed output is sent to a WebSocket server where the NLU and post-processing occurs, and the DM generates appropriate responses. Responses are voiced back to users using Google's Speech Synthesis API \cite{google-speech-synthesis}.

The second module performs NLU on the input, extracting and recognizing users' intents based on utterances provided by the VUI. The NLU module is syntactically-constrained and uses a regex expression-based semantic parser to determine intent and extract semantic information. %
In future studies, we will implement customized NL and speech models, and compare this system with our current constrained system to determine usability and cognitive load effects.

The third module is the DM, which is responsible for keeping track of the conversation, goals between users and the system, and the agent state. Agent states include program creation, editing and execution states, and are managed by a finite-state machine. %

Given information extracted by the NLU, the DM creates a ``user goal.'' The goal contains the user's intent (e.g. intent to create a variable) and necessary actions the system must take to complete the goal. These actions are referred to as ``agent goals.'' For example, when a user wants to create a variable, the DM will ask for a name and an initial value if they weren't originally provided by the user. In this case, the user goal is to create a variable, while the agent goals are to ask for the name and initial value. The latter behavior is commonly known as slot-filling. %
Completing agent goals leads to various possible changes, including state changes and new additions to the program. The agent goals and state determine the specific response \textsc{Convo} provides after a given user command.

The fourth module is the PE, which is responsible for program-related tasks like editing and execution. The PE interacts with the DM, receiving program-related commands and actions and returning program context and state. During program creation, the editor keeps a program representation in memory. The representation is a list of actions that the agent performs when executing the program, and can be exported to other formats (e.g. JSON, Javascript, Python). During editing, the PE keeps track of the program state, which contains information such as defined variables and the current action.

\section{User Study}\label{sec:study}
We conducted a user study to evaluate the effectiveness of \textsc{Convo} and to understand the user needs of a conversational programming environment.

\subsection{Participants}
We recruited 45 participants through university mailing lists and flyers, with 27 males, 17 females, and 1 unspecified. Participants ranged from local high school students to members of local universities to community members from elsewhere. The minimum age of participants was 14, maximum was 64, and mean 25.3. Based on survey results, 12 users self-identified as ``novice'' (users with little to no programming experience) and 33 users self-identified as ``advanced'' (users who had completed a programming course or had experience in object-oriented programming). All 45 participants completed at least one part of the user study. Participants were given a \$20 or \$30 Amazon gift card depending on whether they identified as novice or advanced.

\subsection{Procedure}
Upon arriving to the user study, participants were presented with an informed consent form, in which they agreed to be audio- and keystroke-recorded. Participants were asked to provide their own laptops, but were provided with earbuds for the task. Before starting the study, participants read detailed instructions about what to expect during the study and watched a video on how to use the text-input, voice-input, and voice-or-text systems, and filled out a demographics questionnaire. 

Participants interacted with the \textsc{Convo} programming environment in three stages: the practice stage, the novice stage, and the advanced stage, with the advanced stage for advanced participants only. The practice stage was designed for participants to familiarize themselves with the environment. At each stage, participants interacted with three systems. Each system had a goal for the user to complete. Participants were shown what success looked like for each goal through a video prior to starting. Participants could not move on to the next goal until they successfully completed the current goal. After interacting with a system in a particular stage, participants filled out a questionnaire about their experience.

We performed a mixed between- and within-subject test, where the between-subject conditions were the novice and advanced stages and the within-subject condition was input modality type. We randomized the order of the systems and introduced slight variations to the goals to account for learning effects. Participants were given as much time as they needed to complete all tasks, and could raise their hand to ask for help if they had questions. Questions were addressed following a strict protocol to ensure participants received the same advice for specific technical issues. After completing the study, participants filled out a final questionnaire.

\subsection{Study Tasks}
\begin{figure}[ht]
    \includegraphics[width=\linewidth]{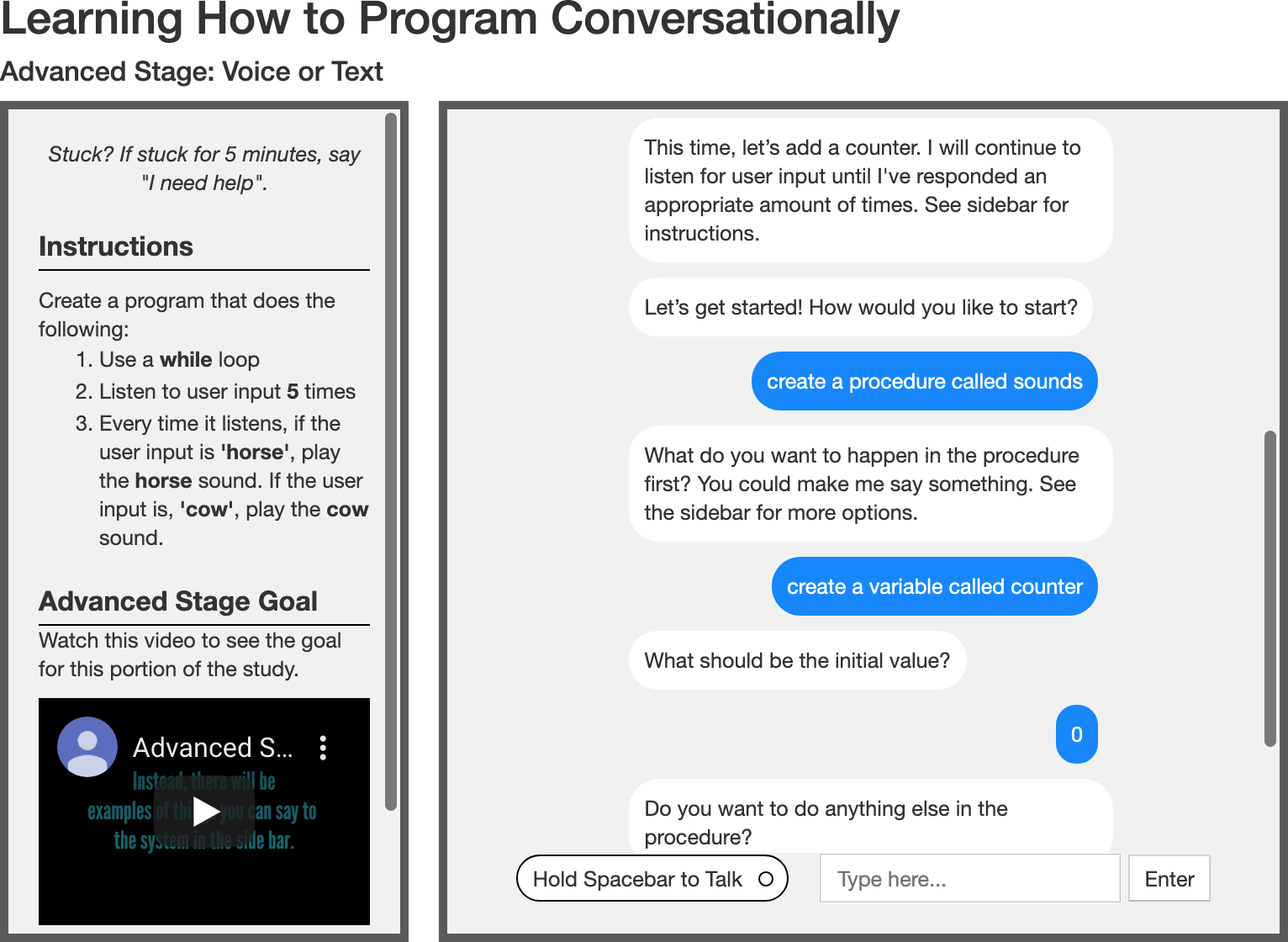}
    \caption{The advanced stage of the study using the voice-or-text-based system, which shows both the record button and text box %
    for input.}
    \label{fig:userinterface}
\end{figure}

At every practice, novice, and advanced stage, participants were given three tasks, or goals, to complete, one for each system (voice-input, text-input and voice-or-text systems). The goals varied slightly but were similar enough that participants would produce similar actions while interacting with different systems. Goals were randomly matched to systems. The set of tasks participants were asked to complete were:

\begin{enumerate}
    \item Practice Stage: Create a program where \textsc{Convo} says ``hello world" in audio format. We varied the phrase \textsc{Convo} should say three times.
    \item Novice Stage: Create a program where \textsc{Convo} listens for user input and plays two different animal sounds (e.g. If I say ``cat", play ``meow"). We varied the type of animal sounds required three times.
    \item Advanced Stage: Create a program where \textsc{Convo} continuously listens for user input for a set number of times and plays the corresponding animal sound. We varied the number of times \textsc{Convo} listens for user input and the required animal sounds three times. Only advanced participants completed the Advanced Stage.
\end{enumerate}
The novice goals required participants to use conditionals and variables. The advanced goals required participants to build off of novice goals and use a loop to generate the specified animal sounds multiple times. The user interface for the study is shown in Fig. \ref{fig:userinterface}.

\subsection{Data Collection}
Each participant completed a questionnaire about their demographic and programming background. All text- and voice-input transcripts were recorded throughout the study. We also recorded the number of times participants asked for help and the number of resets they were given when they got stuck and wanted to redo a particular goal. After every completed task, we collected the following measures:
\begin{itemize}
    \item Time: The total time duration for a participant to complete a goal.
    \item Usability: Participants responded to \textit{``I found it difficult to complete the goal with the [input]-based system."} and \textit{``I found programming with the [input]-based system difficult to use."} on a 5-point Likert scale.
    \item Satisfaction: Participants responded to \textit{``I am satisfied programming with the [input]-based system."} on a 5-point Likert scale.
    \item Efficiency: Participants responded to \textit{``I found programming with the [input]-based system efficient to use."} on a 5-point Likert scale.
    \item Preferences: Participants answered free-form questions about what they liked and disliked in using the system.
    \item Desired features: Participants answered free-form questions about what features they wished to add.
\end{itemize}

At the conclusion of the study, users filled out a questionnaire with 5-point Likert scale questions comparing the three systems and free-form questions 
asking participants what they would ask the conversational agent, challenges they ran into, and questions they had about the system. 

We used analysis of variance (ANOVA) for between-subjects analyses and repeated measures ANOVA for within-subjects analyses.

\subsection{Results}\label{sec:results}

\subsubsection{Quantitative Analysis}\label{sec:quant}

The type of input modality had a significant effect on participants' perception of the system. Our results show that both novice and advanced participants strongly preferred the text-based system over the voice-based system. Participants felt it was more difficult to complete the programming goals with the voice-based system, and were generally more satisfied with the text-based system.

Novice participants made significantly more incorrect utterances with the voice-based system ($M$=17.38) compared to the text-based ($M$=1.38, $F_{1,38}$=17.79, $p$=0.0001) and voice-or-text-based systems ($M$=4.77, $F_{1,38}$=11.39, $p$=0.0017), whereas no significant difference was observed for advanced participants. In addition, novice participants were more satisfied with the voice-or-text based system ($M$=2.61) than the voice-based system ($M$=3.44, $F_{1, 35}$=15.90, $p$=0.0003), and found the voice-or-text-based system ($M$=2.66) more efficient to use than the voice-based system ($M$=3.47, $F_{1, 35}$=14.18, $p$=0.0006). There was no significant difference in preference observed by advanced participants. 

Advanced participants perceived the voice-or-text-based system ($M$=2.94) to be more difficult to use compared to the text-based system ($M$=3.5, $F_{1,15}$=6.36, $p$=0.02); there was no significant difference found for novice participants. Overall, novice participants found the voice interaction of the voice-based and voice-or-text-based systems to be useful and enjoyable, whereas advanced participants tended to disagree more with those statements (see Fig. \ref{fig:likert-advanced}).

Prior programming knowledge and gender did not have a significant effect on completion time for all participants. Novice participants and advanced participants completed the practice and novice stages in around the same time. There was also no significant difference between the number of voice utterances and text utterances during the novice stage. Advanced participants tended to use more text utterances than voice utterances during the advanced stage.

To investigate cognitive load effects, we examined the number of resets of the system (as participants mentioned they reset due to forgetting where they were in the program they were creating), time to goal completion, and number of times users asked for help. Note that we only analyzed the advanced stage for cognitive load, since the instructions were provided line-by-line in the novice stage (i.e., minimal cognitive load involved), whereas users needed to %
determine which steps to take next on their own in the advanced stage (i.e., significant cognitive load involved).
There was no observed significant difference in the number of times asked for help with the voice-input, text-input, and voice-or-text system. The input modality also did not have a significant effect on the number of resets or time to goal completion during the advanced stage. %

\subsubsection{Qualitative Analysis}\label{sec:qual}
\begin{figure}[hbt]
    \includegraphics[width=\linewidth]{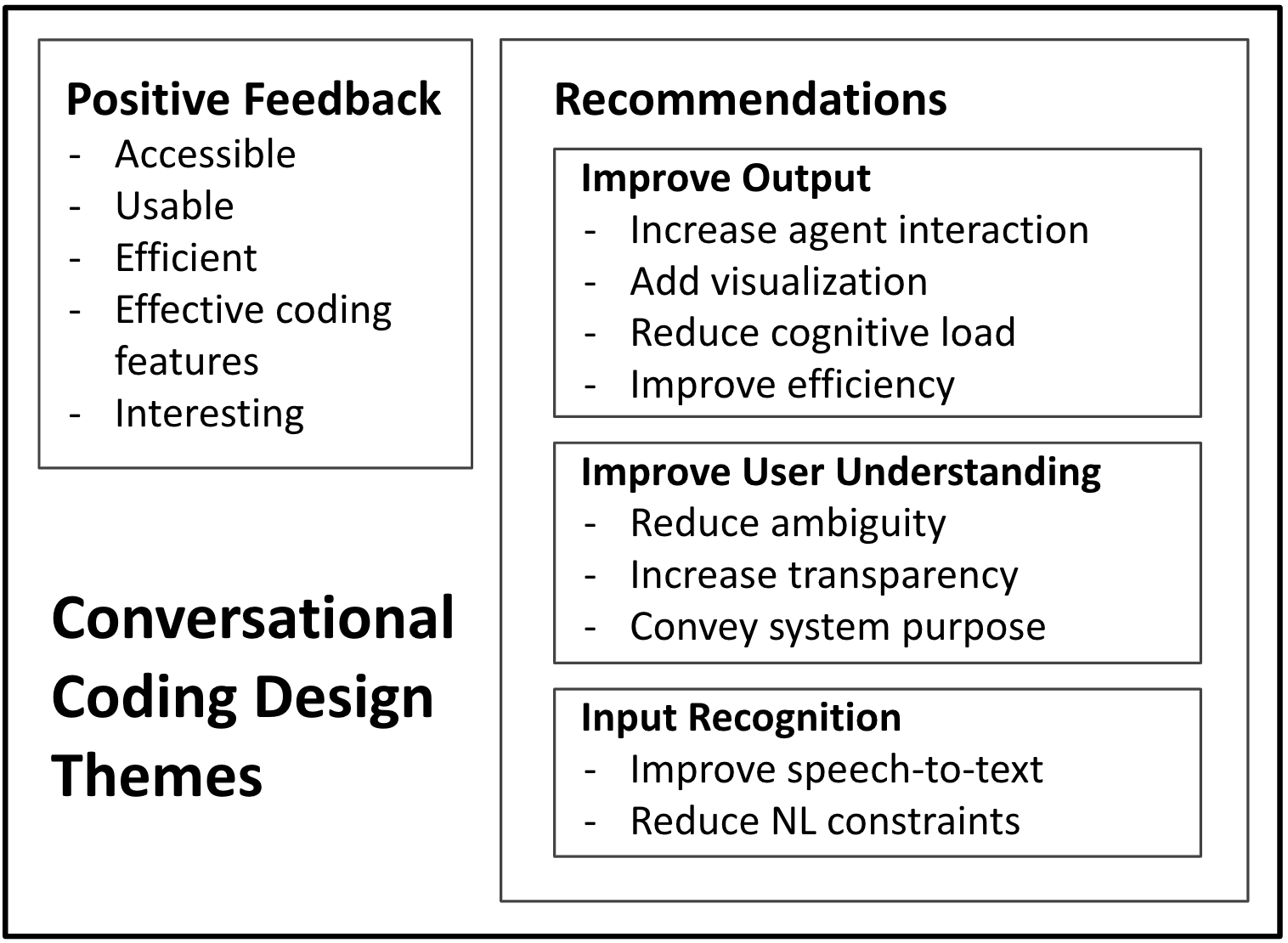}
    \caption{The theme hierarchy created during open coding.}
    \label{fig:themes}
\end{figure}

We organized the free-form responses and analyzed for patterns using an inductive approach \cite{inductive-approach} (i.e. open coding). We identified fourteen design themes, which fell into two main categories, \textit{positive feedback} and \textit{recommendations} (see Fig. \ref{fig:themes}).

We coded 651 occurrences of these themes and show representative quotations below. Themes for the \textit{positive feedback} category follow:

\textbf{Efficient} (49/651): ``I liked how quick it was. Having to just speak to program is far quicker than typing [...]'' %

\textbf{Usable} (48/651): ``I liked how straight-forward and logical it is because it translates the logic of the code into everyday speak.'' %

\textbf{Accessible} (32/651): ``I liked the availability of the text option because it usually would take me a few attempts to get the voice working.'' %

\textbf{Effective coding features} (9/651): ``I liked that it tried to catch cases like `not having a false condition'. I imagine this will be super useful in recommending base cases for recursion problems'' %

\textbf{Interesting} (6/651): ``It feels cool to do this - I can imagine coding while driving or doing housework.'' %

\medskip
Themes for the \textit{recommendations} category regarding improving the agent's output follow:

\textbf{Increase agent interaction} (91/651): ``[I would add] a spellchecker, like if a word is spelled incorrectly it could say `You said `dune', did you mean `done'?''

\textbf{Add visualization} (72/651): ``[I would add] some sort of visualization of the function being built up as interaction progresses''

\textbf{Improve efficiency} (30/651): ``It also seems quite inefficient to figure out the right way to express a statement in actual words that otherwise can be typed in a programming language [...]'' %

\textbf{Reduce cognitive load} (12/651): ``I can't see my program and I have to remember what's going on, that will become infeasible very quickly.'' %

\medskip
Examples from the \textit{recommendation} category regarding improving users' understanding follow:

\textbf{Increase transparency} (25/651): ``[I would ask] `How do you recognize the voices?
Do you use any sort o [\textit{sic}] machine learning to recognize the accents?' ''

\textbf{Reduce ambiguity} (12/651): ``I'm interested in how does the program differentiate similar commands.'' %

\textbf{Convey system purpose} (9/651): ``Who is the intended audience and what sort of programs do you imagine them writing? [...]'' %

\medskip
Examples from the \textit{recommendation} category regarding improving the agent's recognition follow:

\textbf{Improve speech-to-text} (190/651): ``Differentiating between voices and then telling the difference with accents [was a challenge for the system]'' %

\textbf{Reduce NL constraints} (66/651): ``[...] Allowing more variability in what I can say to the agent to get it to do the same command would feel more natural.'' %

\begin{figure}[ht]
    \includegraphics[width=\linewidth]{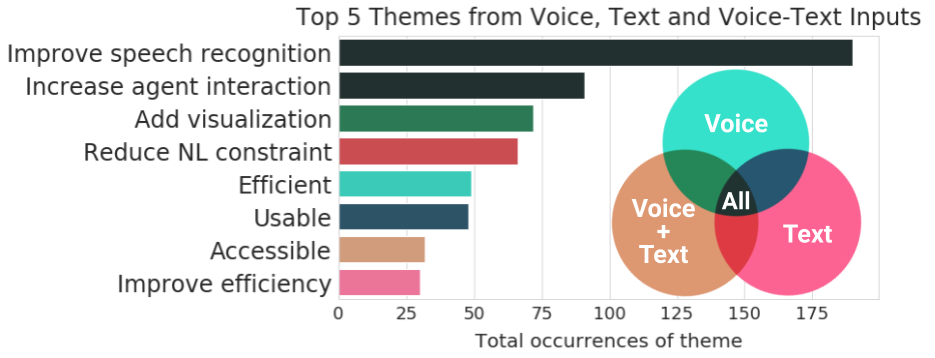}
    \caption{Total number of occurrences for the top five themes from each system survey. The voice-input system responses emphasized efficiency; text-input, a need to improve efficiency; and voice-or-text, accessibility. Note how the bars' colors represent which input system(s) the theme came from (e.g., pink represents text-input system, and green represents voice-input and voice-or-text systems).  %
    }
    \label{fig:topthemes-vtsystems}
\end{figure}

\medskip
As shown in Fig. \ref{fig:topthemes-vtsystems}, six of the top seven themes for novice and advanced users were the same, including \textit{improve speech recognition}, \textit{increase agent interaction}, and \textit{add visualization}. Novice users emphasized increasing transparency over efficiency, and vice versa for advanced users.

\begin{figure}[ht]
    \includegraphics[width=\linewidth]{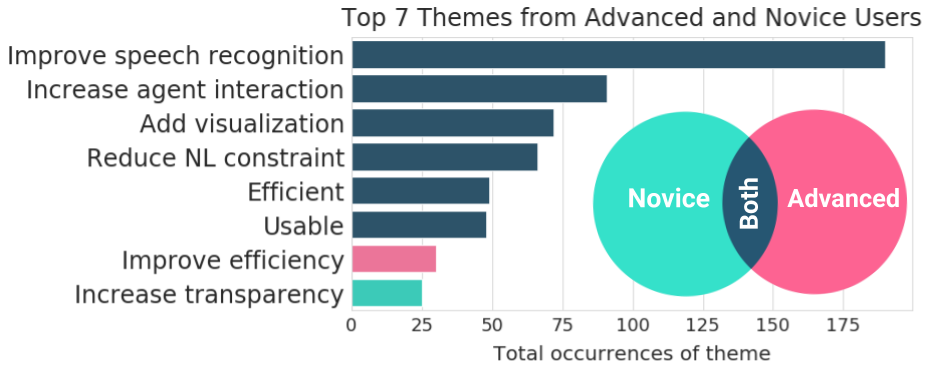}
    \caption{Total number of occurrences for the top seven themes from advanced user responses and top seven from novice user responses. Novice responses emphasized transparency over efficiency. Note how the colors represent which user group(s) the theme came from (e.g., pink represents a top theme from novice users, dark blue represents a top theme from both novice and advanced users).}
    \label{fig:topthemes-userlevel}
\end{figure}

Among input modalities, participants emphasized \textit{improve speech recognition} and \textit{increase agent interaction} for all three systems (voice-input, text-input, and voice-or-text) (see Fig. \ref{fig:topthemes-vtsystems}). The voice-input system responses emphasized efficiency; the text-input system responses emphasized improving efficiency; and the voice-or-text system responses emphasized accessibility. Both the voice- and text-input system responses emphasized usability; both the voice-input and voice-or-text system responses emphasized adding a visualization; and both the text-input and voice-or-text systems emphasized reducing the NL constraint.

\section{Design Recommendations}\label{sec:design-rec}
Through the quantitative and qualitative analyses, we identified  six main design recommendations for future conversational programming systems.

\begin{figure}[ht]
    \includegraphics[width=\linewidth]{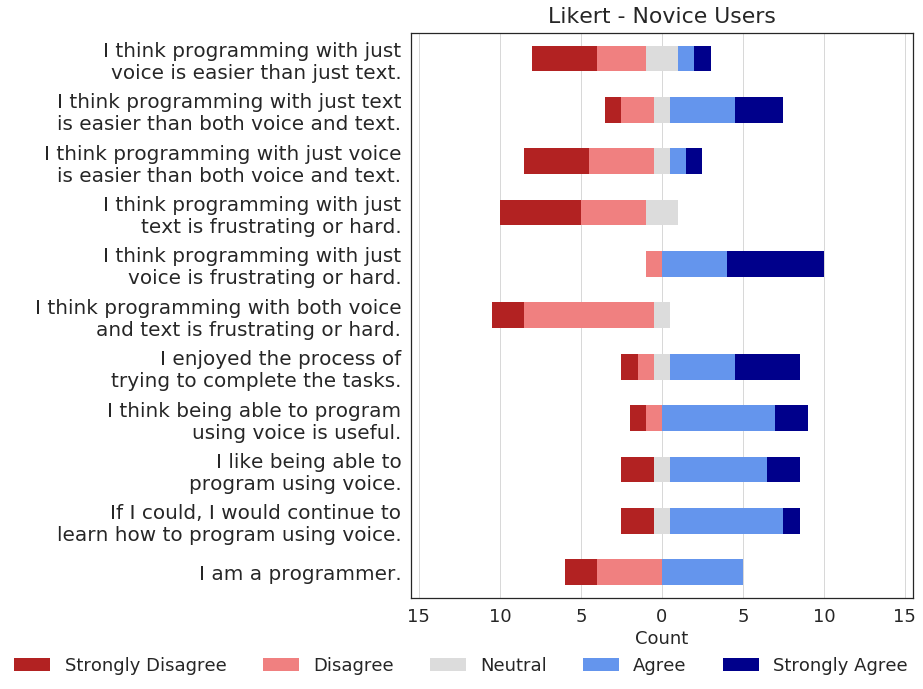}
    \caption{Novice user responses to Likert scale questions. Novices generally found voice useful and enjoyable.}
    \label{fig:likert-novice}
\end{figure}

\begin{figure}[ht]
    \includegraphics[width=\linewidth]{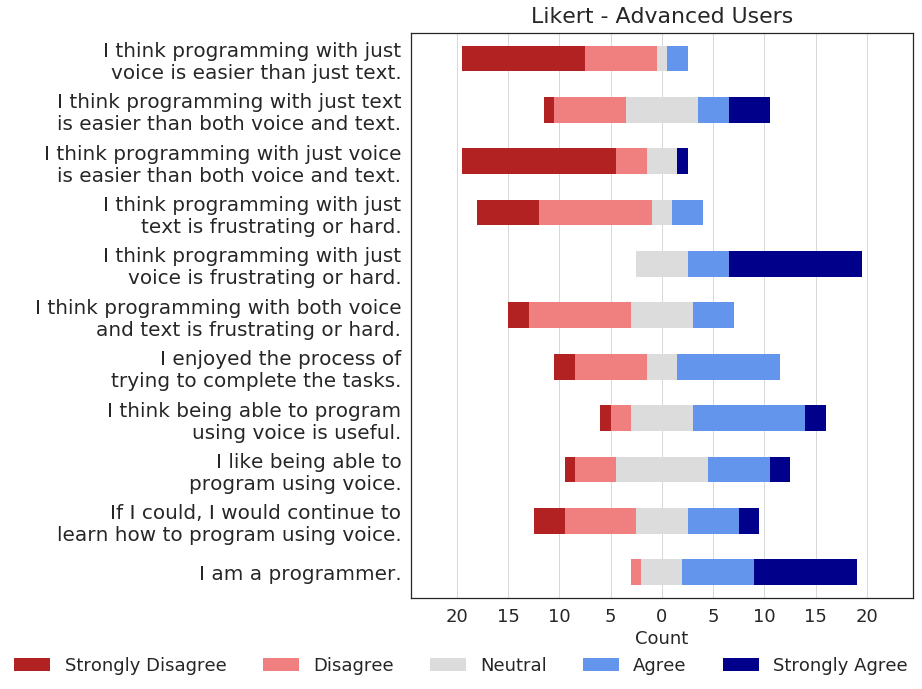}
    \caption{Advanced user responses to Likert scale questions. Advanced user responses tended to be less favorable towards voice than novice responses.}
    \label{fig:likert-advanced}
\end{figure}

\textbf{Tailor to programming experience and task.}
Our results suggest that conversational programming systems should be tailored to their audiences due to differences in user preferences and abilities. We found that novice users generally found voice-input useful and enjoyable, whereas advanced users tended to view it less so, as shown in  Fig. \ref{fig:likert-novice} and \ref{fig:likert-advanced}. Furthermore, although there was no significant difference between the overall number of voice- and text-inputs, in the advanced stage users tended to type rather than speak ($p$=0.003). Advanced users also perceived voice-or-text to be more difficult than text ($p$=0.02), but there was no significant difference for novice users. Thus, for an advanced audience, it may be more important to have a text-input option than for a novice audience, and for an introductory audience, a voice-input system may be more useful than for an advanced one. %

We also found that some advanced users found NL programming cumbersome, likely because they were used to syntax-restricted programming languages (e.g., ``It also seems quite inefficient to figure out the right way to express a statement in actual words that otherwise can be typed in a programming language using very specialized characters.''), whereas novice users tended to praise the naturalness of the language (e.g., ``I liked the simplicity of using the normal talk, as in not coding necessarily''). This is further reflected in how ``improve efficiency'' was found in advanced users' the top seven themes, but not novice users' (see Fig. \ref{fig:topthemes-userlevel}). Thus, NL may be a better fit for an educational, introductory tool than an advanced tool.

\textbf{Design a flexible, accessible system.}
Our results suggest conversational programming systems should be accessible through both voice- and text-input. Participants found value in both modalities, often citing voice as efficient (see Fig. \ref{fig:topthemes-vtsystems}) and text as accurate. Many participants had comments similar to, ``I liked being able to use the voice for longer commands, and the text for shorter commands or misunderstood commands''. This was supported by the significant difference in number of characters ($p$=0.004) and words ($p$=0.003) per voice utterance over text utterance (i.e., longer voice utterances). Furthermore, when using the voice-or-text system, participants used both voice and text input, and there was no statistical evidence for a difference in how many times participants spoke versus typed.

From an accessibility standpoint, it makes sense to provide both input options, and allow each of them to stand alone (such that the system is completely accessible by voice-only and text-only). With current technologies, however, this may be difficult to achieve. The Google Cloud Speech-to-Text \cite{google-cloud-speech-to-text} ASR system we used---which is often recognized as the gold standard \cite{best-asr-google,best-asr-google2}---did not seem sufficient for programming. Many participants commented on this (e.g., ``Sometimes it had problems understanding my speech, so I resorted to typing things.'', ``It seems like if speech recognition worked well, it would be a better choice, but having this [text option] is useful'') and we found that the most common theme was to \textit{improve speech recognition}. Thus, until speech recognition systems improve, it may be infeasible to have a standalone voice-input system.

\textbf{Design a transparent system.}
Many participants described how they would appreciate being able to ask the system how it works. Some questions included:
\begin{itemize}
    \item ``What kind of nueral [\textit{sic}] network do you run on?''
    \item ``How do you understand what I’m saying?''
    \item ``How do you map my phrases to commands?''
    \item ``What kind of voice recognition is used?''
    \item ``Why didn't the agent understand me?''
    \item ``How do you register what I'm saying? Should I speak slower/faster? How can I make it easier for you to understand me?''
    \item ``Do you use any sort o [\textit{sic}] machine learning to recognize the accents?''
\end{itemize}

Transparency was one of the top occurring themes for novice users (see Fig. \ref{fig:topthemes-userlevel}) and especially important when developing AI systems for education.

\textbf{Design with visualizations.}
A common theme in the free-form responses was the desire for code visualizations. This was in the top seven commonly occurring themes for both novice and advanced users, and the top five themes for both the voice- and text-based systems. Specifically, users asked for ways to ``visualize where [they] are in the program'', view a ``representation of the code [they were] making'', ``see [...] variable names or the name of the procedure'', see ``the current state of the program, or at least [...] which level [they]'re at'', and visually ``modify [their] previous lines that were misinterpreted''. This makes sense, as current technology focuses heavily on visual systems and computer screens, and voice-only systems can force high memorization requirements on users. %
Nonetheless, depending on a system's intended audience, one may choose to avoid visualizations or make them non-essential to the system for accessibility reasons. %

\textbf{Design to reduce cognitive load.}
In the thematic analysis, some participants mentioned high cognitive load due to a lack of visualizations (e.g., ``I found it quite challenging to figure out the logic of the program entirely in my head; [...] %
it felt like I had to figure it all out before entering anything.''%
). In future studies, we will analyze cognitive load effects of integrating visualizations into \textsc{Convo}. We expect this will reduce the cognitive load for sighted users. Other design features to potentially reduce cognitive load include decreasing the constraint on the NL input such that users will no longer have to remember specific phrases, and improving the speech recognition model such that people don't have to repeat phrases as often, and are more likely to remember where they are in the program.

For all cognitive load indicators (number of resets of the system, time to goal completion, and number of times users asked for help), we found no evidence for a significant difference between the voice-based, text-based, and voice-or-text-based systems; %
thus, voice-based, text-based and voice-or-text-based systems may be viable options when designing for cognitive load.

\textbf{Improve ASR and NLU.}
The most common theme in the free-form responses was to improve speech recognition. %
As mentioned previously, we used the Google Cloud Speech-to-Text \cite{google-cloud-speech-to-text} ASR system---which is often recognized as the top online ASR \cite{best-asr-google,best-asr-google2}---for \textsc{Convo}. Evidently, current ASR systems are not sufficient for fully standalone voice-based, NL programming systems. One potential avenue for improvement is to develop a custom NL programming ASR model that incorporates common NL programming phrases, like ``create a variable'', to ensure recognition of those phrases. Nonetheless, by training on specific phrases, this may cause the model to be less robust to new phrases, which would somewhat defeat the purpose of a generalizable NL system.

In addition to improved speech recognition, participants desired reduced constraint on NL input (e.g., ``It's a very cool idea, and with expanding the dictionary it could work better.'', ``I expect more natural-language input support such as `nope', `no thanks', etc. would be valuable as well.''). Reducing NL constraint was a top theme in both the text-based and voice-or-text-based systems, as well as in both novice and advanced users' responses (see Fig. \ref{fig:topthemes-userlevel} and \ref{fig:topthemes-vtsystems}).

We are currently developing an unconstrained NL version of \textsc{Convo} to understand whether this improves or reduces performance, as there has been research questioning the suitability of unconstrained NL for programming \cite{hci-handbook,unconstrained-nl-bad}. Nevertheless, with additional ambiguity reduction techniques, such as conversational QA and immediate feedback from the agent, unconstrained NL may become suitable for introductory, educational NL programming, especially due to the positive feedback in this area from the free-form responses (e.g., ``It gives feedback, which is really useful'', ``The process is pretty interactive and fun. The idea of using natural language to code is great and the system reacts very fast.'', ``Feedback is immediate.'').

\section{Conclusions}\label{sec:conclusions}
In this study, we investigated the effectiveness of voice-based, text-based, and voice-or-text-based systems in a conversational programming environment. We analyzed the systems in terms of difficulty, efficiency, and cognitive load indicators through free-form responses, Likert scale questions, and user activity during programming task completion. Our results show a desire for and optimism about conversational programming, especially in introductory programming systems. Future conversational and interactive ML systems should consider the following six design recommendations: (1) Tailor to programming experience and task, (2) Design a flexible, accessible system, (3) Design a transparent system,  (4) Design with visualizations, (5) Design to reduce cognitive load, and (6) Improve ASR and NLU. %
Future iterations of \textsc{Convo} will include addressing questions about the effects of visualizations and reducing NL constraints in terms of usability and cognitive load, and the effectiveness of conversational programming for learning computational thinking skills and taking computational action. 

\section{Acknowledgements}
We would like to thank Hal Abelson, Marisol Diaz, Selim Tezel, and Ilaria Liccardi for their support, as well as the participants in our study for their time.

\bibliographystyle{ieeetr}
\bibliography{biblio}

\end{document}